\documentclass[aps,preprint,showpacs,superscriptaddress,groupedaddress]{revtex4} 

\usepackage[dvips]{graphicx}
\usepackage{dcolumn}   
\usepackage{bm}        
\usepackage{amssymb}   
\usepackage[utf8]{inputenc}
\usepackage{mathrsfs}
\usepackage{amsmath}

\usepackage[english]{babel}

\usepackage[colorlinks]{hyperref}

\hypersetup{colorlinks=true,citecolor=blue,linkcolor=blue,urlcolor=blue}
\usepackage[usenames]{color}
\usepackage{cancel}

\usepackage{ulem}
\usepackage{orcidlink}
\usepackage{subfigure}

\makeatletter
\newcommand{\linethrough}{\mathpalette\@thickbar}
\newcommand{\@thickbar}[2]{{#1\mkern0mu\vbox{
    \sbox\z@{$#1#2\mkern-1.5mu$}%
    \dimen@=\dimexpr\ht\tw@-\ht\z@+2\p@\relax 
    \hrule\@height0.5\p@ 
    \vskip\dimen@
    \box\z@}}
}
\makeatother

\begin{document}

\preprint{\small Chaos, Solitons and Fractals \textbf{194} (2025) 116170
\ [\href{https://doi.org/10.1016/j.chaos.2025.116170}{DOI: 10.1016/j.chaos.2025.116170}]
}

\widetext

\title{Long-range interaction of kinks in higher-order polynomial models}

\author{Ekaterina Belendryasova\,\orcidlink{0000-0002-1353-1439}}
\email{bel.eg@ibrae.ac.ru}
\affiliation{The Nuclear Safety Institute of the Russian Academy of Sciences, Moscow 115191, Russia }

\author{Petr A. Blinov\,\orcidlink{0000-0003-4055-6081}}
\email{blinov.pa@phystech.edu}
\affiliation{Moscow Institute of Physics and Technology,
Dolgoprudny, Moscow Region 141700, Russia}

\author{Tatiana~V.~Gani\,\orcidlink{0000-0003-2993-5822}}
\email{gani.t@bk.ru}
\affiliation{M.~V.~Lomonosov Moscow State University, Moscow 119991, Russia
}

\author{Alexander~A.~Malnev\,\orcidlink{0000-0003-4001-1970}}
\email{malnev.aa@phystech.edu}
\affiliation{Moscow Institute of Physics and Technology,
Dolgoprudny, Moscow Region 141700, Russia}

\author{Vakhid A. Gani\,\orcidlink{0000-0001-5825-5696}}
\email{vagani@mephi.ru}
\affiliation{National Research Nuclear University MEPhI\\ (Moscow Engineering Physics Institute), Moscow 115409, Russia}
\affiliation{
National Research Centre ``Kurchatov Institute'',
Moscow 123182, Russia}

\begin{abstract}

We obtain asymptotic estimates of the interaction forces between kink and antikink in a family of field-theoretic models with two vacua in (1+1)-dimensional space-time. In our study we consider a new class of soliton solutions previously found in our paper [Chaos, Solitons and Fractals 165 (2022) 112805]. We focus on the case of kinks having one exponential and one power-law asymptotics. We show that if the kink and antikink are faced each other with long-range tails, the force of attraction between them at large separations demonstrates a power-law decay with the distance. We also performed numerical simulations to measure the interaction force and obtained good agreement between the experimental values and theoretical estimates.

\end{abstract}



\maketitle






\section{Introduction}
\label{sec:Introduction}

In this paper we study low-dimensional topological solitons --- kink solutions of the nonlinear Klein--Gordon equation with a nonlinearity of some rather general form \cite{Manton.book.2004,Shnir.book.2018}. The above equation can be obtained as the equation of motion for a real scalar field in $(1+1)$-dimensional space-time from Lagrangian field theory. The subject of the study is the interaction force between a kink and an antikink separated by a large distance. Moreover, we are interested in the case of power-law asymptotic behavior of kinks. The interaction force arises in non-integrable models as a result of the fact that, due to nonlinearity, kink and antikink located at a finite distance from each other are not an exact solution to the equation of motion. Note that there are exact two-soliton solutions in the integrable sine-Gordon model \cite{Kevrekidis.book.2014} and in this case the interaction force does not arise.

Kinks occur in many physical contexts. It seems completely impossible to make an exhaustive review here, so we mention only a few cases. It was shown \cite{Buijnsters.PRL.2014} that in thin membranes of some magnetic materials, such as yttrium iron garnet $Y_3Fe_2(Fe O_4)_3$, domains change their sizes due to the movement of boundaries by means of kink deformations. Running and interacting kink deformations observed in a graphene nanoribbon --- a narrow long sample of a two-dimensional material compressed from the sides \cite{Yamaletdinov.PRB.2017}. The kinks observed in \cite{Yamaletdinov.PRB.2017} were well fitted by the $\varphi^4$ kinks, however, some observed asymmetry indicated the need to use other models, e.g., $\varphi^6$ or $\varphi^8$. For a nematic liquid crystal 4-n-methoxybenzylidene-n-butylaniline placed between glass substrates with a conducting coating of $SnO_2$, the collision of dislocations in an extended defect was qualitatively well described by a two-soliton solution \cite{Delev.JETP_Lett.2019_1,Delev.JETP_Lett.2019_2}.

Until recently, only kinks with exponential asymptotics were studied. Perhaps the most famous example is the kink of the $\varphi^4$ model. The history of numerical and analytical study of the $\varphi^4$ kink interactions begins in the 1970s, when the works of A.~E.~Kudryavtsev \cite{Kudryavtsev.JETPLett.1975}, B.~S.~Getmanov \cite{Getmanov.JETPLett.1976}, and S.~Aubry \cite{Aubry.JChemPhys.1976} appeared. More detailed studies of the collision behavior of these kinks were carried out several years later by M.~J.~Ablowitz, M.~D.~Kruskal, J.~F.~Ladik \cite{Ablowitz.SIAM_JAM.1979} and by T.~Sugiyama \cite{Sugiyama.PTP.1979}. Later, other researchers obtained impressive results, in particular D.~Campbell and co-authors \cite{Campbell.PhysD.1983.phi4}. See also review \cite{Belova.UFN.1997} and book \cite{Kevrekidis.book.2019}.

In the last decade, attention to the interaction of kinks with power-law tails has been gradually increasing. In this case, the interaction decreases with separation in a power-law manner, i.e., kink and antikink ``feel'' each other at very large distances. The interaction was studied numerically and analytically using approximate methods \cite{Radomskiy.JPCS.2017,Christov.PRL.2019,Manton.JPA.2019,dOrnellas.JPC.2020,Campos.PLB.2021,Belendryasova.CNSNS.2019,Christov.PRD.2019,Christov.CNSNS.2021,Campos.JHEP.2024}. Analytical estimates of the force were obtained in \cite{Radomskiy.JPCS.2017, Christov.PRL.2019, Manton.JPA.2019, dOrnellas.JPC.2020, Campos.PLB.2021}, see also \cite{Khare.JPA.2019,Khare.FP.2022}. For numerical modeling of kink-antikink interactions in the power-law case, special methods for construction of initial configurations have been developed \cite{Christov.PRD.2019,Christov.CNSNS.2021,Campos.JHEP.2024}.

In a recent paper \cite{Blinov.CSF.2022}, kink solutions were obtained in a family of field-theoretic models with polynomial potentials having two true vacua. In addition, the asymptotic properties of the resulting kinks were investigated. This paved the way for studying the problem of kink-antikink forces. In this paper, we focus on obtaining asymptotic estimates of these forces in the case of large distances between the solitons.

Our paper is organized as follows. In Section~\ref{sec:Model} we briefly describe the field-theoretic model under consideration, present kink solutions and their main properties required for further presentation. Asymptotic estimates of kink-antikink forces are obtained in Section~\ref{sec:Forces}. In this section we also compare our results with those obtained previously in close models or in special cases. In Section \ref{sec:Numerical} we present the results of numerical simulations that allowed us to obtain experimental values of the kink-antikink force and compare them with theoretical predictions. Finally, Section~\ref{sec:Conclusion} presents a brief discussion, concluding remarks and prospects for future work. For completeness, the Appendix \ref{ap:num-details} provides technical details of the numerical simulation. In particular, we explain (i) how the initial conditions for kinks with power-law tails were generated, (ii) how the problem of the evolving initial kink+antikink configuration was solved numerically, (iii) how the force of the kink-antikink interaction was found.

\section{Field-theoretic model and kink solutions}
\label{sec:Model}

We consider a Lorentz-invariant field-theoretic model with a real scalar field in $(1+1)$ dimensions, defined by the Lagrangian
\begin{equation}\label{eq:lagrangian}
    \mathcal{L} = \frac{1}{2}\left(\frac{\partial \varphi}{\partial t}\right)^2-\frac{1}{2}\left(\frac{\partial \varphi}{\partial x}\right)^2-V(\varphi)
\end{equation}
with a polynomial potential of the form
\begin{equation}\label{eq:potential_general}
    V(\varphi) = \frac{1}{2}\left(1+\varphi\right)^{2}\left(1-\varphi\right)^{2n},
\end{equation}
where $n\ge 2$ is an integer. The minimum points (zeros) of the potential \eqref{eq:potential_general} are $\bar{\varphi}_1^{}=-1$ and $\bar{\varphi}_2^{}=1$, also called vacua of the model. The Lagrangian \eqref{eq:lagrangian} yields the equation of motion for the field $\varphi(x,t)$:
\begin{equation}\label{eq:eqmo}
    \frac{\partial^2\varphi}{\partial t^2} - \frac{\partial^2\varphi}{\partial x^2} + \frac{dV}{d\varphi} = 0.
\end{equation}
The conserved energy looks like
\begin{equation}\label{eq:energy}
    E[\varphi] = \int_{-\infty}^{\infty}\left[\frac{1}{2} \left( \frac{\partial\varphi}{\partial t} \right)^2+\frac{1}{2} \left( \frac{\partial\varphi}{\partial x} \right) ^2+V(\varphi)\right]dx.
\end{equation}

The static kink (antikink) solution is a strictly monotonically increasing (decreasing) function $\varphi_{\rm K}^{}(x)$ ($\varphi_{\rm\bar{K}}^{}(x)$), which satisfies Eq.~\eqref{eq:eqmo} and is consistent with the boundary conditions $\lim\limits_{x\to-\infty}\varphi_{\rm K}^{}(x)=-1$, $\lim\limits_{x\to+\infty}\varphi_{\rm K}^{}(x)=1$ ($\lim\limits_{x\to-\infty}\varphi_{\rm\bar{K}}^{}(x)=1$, $\lim\limits_{x\to+\infty} \varphi_{\rm\bar{K}}^{}(x)=-1$).

Since the potential \eqref{eq:potential_general} is positive, we can introduce a prepotential (often called the superpotential) --- a smooth function $W(\varphi)$ such that
\begin{equation}\label{eq:dwdfi}
    V(\varphi) = \frac{1}{2}\left(\frac{dW}{d\varphi}\right)^2,
\end{equation}
see, e.g., Chapter 5 in \cite{Manton.book.2004} for details. Then the energy of the static kink (antikink), i.e. its mass, can be written as
\begin{equation}\label{eq:static_energy_bps}
    M_{\rm K}^{} = \big|W[\varphi(+\infty)]-W[\varphi(-\infty)]\big|,
\end{equation}
and the mass of the part of the kink between points $x_1^{}$ and $x_2^{}$ is obviously equal to $\big|W[\varphi(x_2^{})]-W[\varphi(x_1^{})]\big|$.

For the potential \eqref{eq:potential_general} in \cite{Blinov.CSF.2022} kinks were obtained in the implicit form $x=x_{\rm K}^{}(\varphi)$:
\begin{equation}\label{eq:kink_for_m_equals_1}
    x - x_0^{} = \frac{1}{2^n}\ln\frac{1+\varphi}{1-\varphi} + \sum_{j=1}^{n-1}\frac{1}{j\cdot 2^{n-j}}\frac{1}{(1-\varphi)^j}.
\end{equation}
If we require that $\varphi=0$ at $x=0$, then $x_0^{}=-\sum\limits_{j=1}^{n-1}\frac{1}{j\cdot 2^{n-j}}$. We can also agree on another centering such that $x_0^{}=0$. Kink \eqref{eq:kink_for_m_equals_1} has left exponential and right power-law asymptotics:
\begin{equation}\label{eq:kink_for_m_equals_1_asymptotics}
    \varphi_{\rm K}^{}(x) \approx
    \begin{cases}
    -1 + 2\exp\left[2^n (x-x_0^{})\right] \quad \mbox{at} \quad x\to-\infty,\vspace{1mm}\\
    \thinspace\thinspace\thinspace\thinspace 1 - \displaystyle\frac{\left[2(n-1)\right]^{1/(1-n)}}{(x-x_0^{})^{1/(n-1)}} \quad \mbox{at} \quad x\to+\infty.
    \end{cases}
\end{equation}
The prepotential for the potential \eqref{eq:potential_general} can be written as
\begin{equation}\label{eq:superpotential_for_m_equals_1}
    W(\varphi) = - \frac{\left(1-\varphi\right)^{n+1}\left[\left(n+1\right)\varphi+n+3\right]}{(n+1)(n+2)}.
\end{equation}
The overall minus sign of the superpotential \eqref{eq:superpotential_for_m_equals_1} does not affect anything and can be omitted. However, we use this form of the superpotential \eqref{eq:superpotential_for_m_equals_1} because it was written in \cite{Blinov.CSF.2022}, where the kink solutions studied in this paper were originally obtained. The mass of the kink \eqref{eq:kink_for_m_equals_1} is
\begin{equation}\label{eq:mass_for_m_equals_1}
    M_{\rm K}^{} = \frac{2^{n+2}}{(n+1)(n+2)}.
\end{equation}
For example, for $n=2$ we get kink
\begin{equation}\label{eq:kink_for_n_equals_2}
    x = \frac{1}{4}\ln\frac{1+\varphi}{1-\varphi} + \frac{1}{2\left(1-\varphi\right)}
\end{equation}
with mass $4/3$ and the power-law asymptotic behavior at $x\to+\infty$:
\begin{equation}\label{eq:kink_for_n_equals_2_asymptotics}
    \varphi_{\rm K}^{}(x) \approx 1 - \frac{1}{2x}.
\end{equation}

Let us now move on to studying the interaction between the kink and antikink. Due to power-law asymptotics, we expect a power-law dependence of the force on the distance in the case when the field between kinks is close to the vacuum value $\bar{\varphi}_2^{}=1$.

\section{Kink-antikink forces}
\label{sec:Forces}

\begin{figure}[t!]
	\begin{minipage}[h]{0.495\linewidth}
    \center{\includegraphics[width=0.99\textwidth]
    {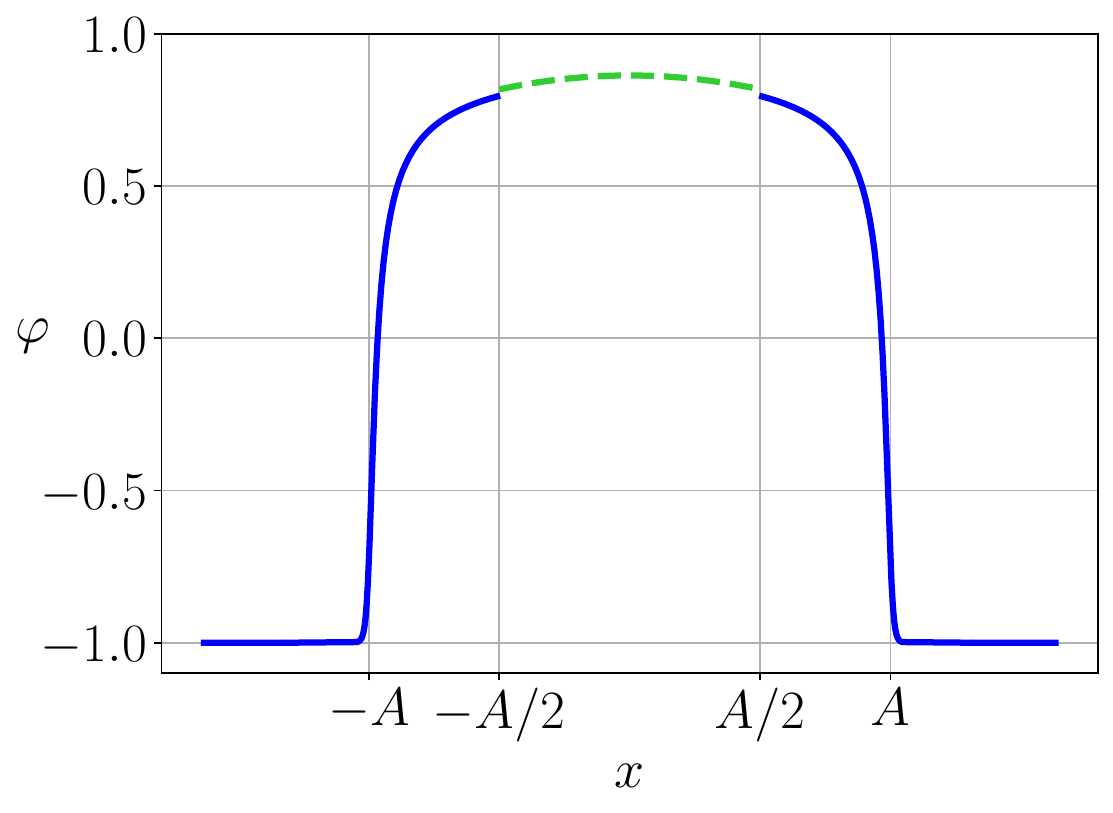}
    \\ (a) }
    \end{minipage}
    \hfill
    \begin{minipage}[h]{0.495\linewidth}
    \center{\includegraphics[width=0.99\textwidth]
    {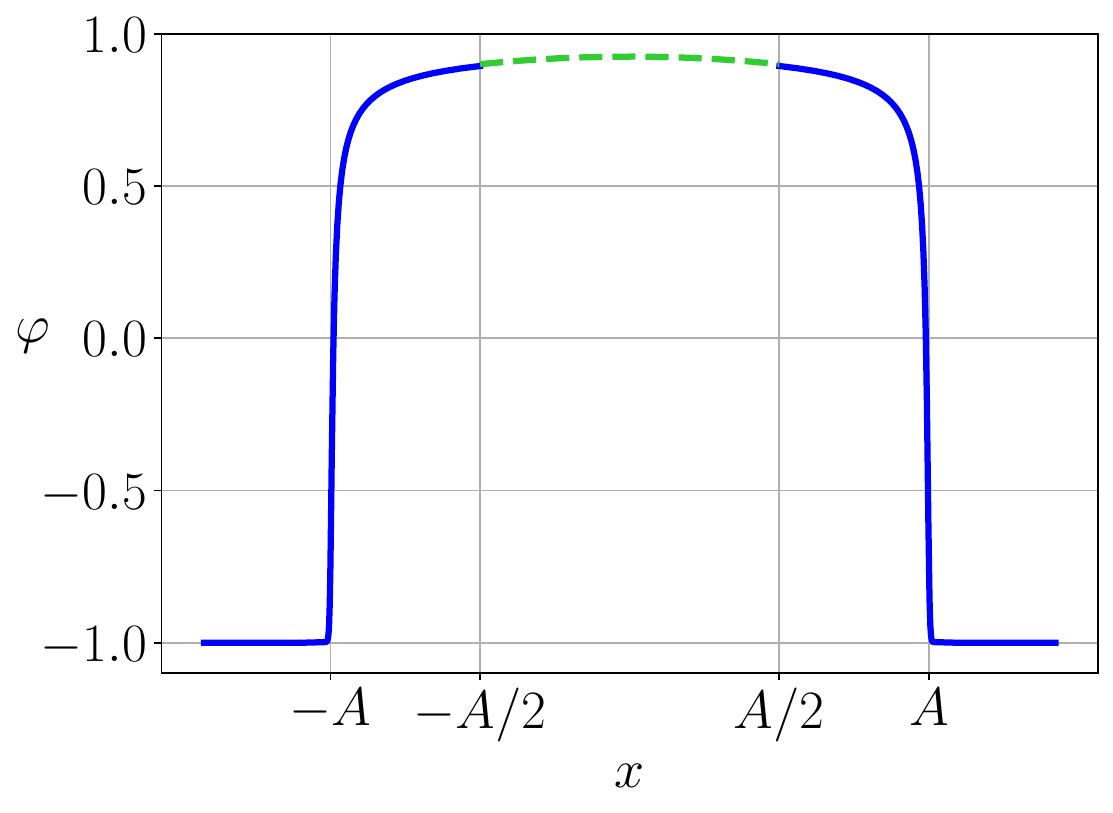}
    \\ (b)}
    \end{minipage}
    \hfill
\caption{Kink-antikink configurations with quadratic interpolation: (a) $n = 3$, $A = 15$; (b) $n = 3$, $A = 50$. The blue solid lines show the exact kinks and antikinks, the green dashed lines is used for the approximating function \eqref{eq:phi_quad}.}
\label{fig:fig1}
    \end{figure}

To calculate the force between a kink and an antikink, we apply a method based on a quadratic approximation of the field between solitons \cite{Manton.JPA.2019} (about calculation kink-antikink force in the case of power-law tails see also \cite{dOrnellas.JPC.2020,Campos.PLB.2021}). Consider static kink and antikink located at points $x=-A$ and $x=A$, respectively, where $A\gg 1$. The kink-antikink configuration we approximate as follows: for $x\le -A/2$ we use an exact kink solution, and for $x\ge A/2$ --- an exact antikink solution. For $-A/2\le x\le A/2$ we approximate the field by a quadratic function of the form
\begin{equation}\label{eq:phi_quad}
    \varphi_{\rm quad}^{}(x) = \alpha + \beta\:x^2.
\end{equation}
The parameters $\alpha$ and $\beta$ are determined from the conditions of continuity of the field and its first derivative at $x=\pm A/2$ (for reasons of symmetry, it is sufficient to use continuity at only one of the two points):
\begin{equation}\label{eq:asymptotics_general}
    \begin{cases}
    \varphi_{\rm quad}^{}(-\frac{A}{2}) = \varphi_{\rm K}^{}(-\frac{A}{2}),\\
    \varphi_{\rm quad}^\prime(-\frac{A}{2}) = \varphi_{\rm K}^\prime(-\frac{A}{2}).
    \end{cases}
\end{equation}
At the point $x=-A/2$, we use approximate asymptotic expression for $\varphi_{\rm K}^{}(x)$ taken from \eqref{eq:kink_for_m_equals_1_asymptotics}. As a result we get
\begin{equation}\label{eq:alpha}
    \alpha = 1 - \frac{1}{\left(n-1\right)^{1/(n-1)}}\left[1-\frac{1}{2(n-1)}\right] \frac{1}{A^{1/(n-1)}},
\end{equation}
\begin{equation}\label{eq:beta}
    \beta = - \frac{2}{\left(n-1\right)^{n/(n-1)}} \cdot \frac{1}{A^{(2n-1)/(n-1)}}.
\end{equation}
The configuration constructed in this way is shown in Fig.~\ref{fig:fig1}.
%
%

To calculate the contribution to the energy from the approximating function \eqref{eq:phi_quad}, we use that $\varphi$ is close to 1, so we can approximately assume that $V(\varphi)\approx 2\left(1-\varphi\right)^{2n}$. Then
\begin{equation}
    E_{\rm quad}^{}(A) \approx \int_{-A/2}^{A/2} \left[ \frac{1}{2}\left(\frac{d\varphi_{\rm quad}^{}}{dx}\right)^2 + 2\left(1-\varphi_{\rm quad}^{}\right)^{2n} \right]dx.
\end{equation}
Substituting \eqref{eq:phi_quad} and taking into account \eqref{eq:alpha} and \eqref{eq:beta}, we obtain
\begin{equation}\label{eq:E_quad}
    E_{\rm quad}^{}(A) \approx \frac{\varepsilon(n)}{A^{(n+1)/(n-1)}},
\end{equation}
where
\begin{equation}
    \varepsilon(n) = \frac{2}{3}\left(n-1\right)^{2n/(1-n)} + 2\left[\left(n-1\right)^{1/(1-n)}-\frac{1}{2}\left(n-1\right)^{n/(1-n)}\right]^{2n}\sum_{k=0}^{2n}\frac{C_{2n}^k}{(2k+1)\left(2n-3\right)^k}.
\end{equation}
Here we use the notation for the binomial coefficient: $C_n^k=\displaystyle\frac{n!}{k!(n-k)!}$.
Then the total energy of the ``kink+antikink'' configuration is
\begin{equation}\label{eq:E_A}
    E(A) = E_{\rm quad}^{}(A) + 2\left[M_{\rm K}^{} - E_{\rm tail}^{}(A)\right],
\end{equation}
where $E_{\rm tail}^{}(A)$ is the mass of the right tail of the kink (i.e.\ the mass of the tail extending into the region $x\ge -A /2$ belonging to the kink centered at $x=-A$), or, which is the same, the mass of the left tail of the kink (i.e.\ the mass of the tail extending into the region $x\le A /2$ belonging to the kink centered at $x=A$):
\begin{equation}\label{eq:E_tail}
    E_{\rm tail}^{}(A) = W(1) - W(\varphi_{\rm K}^{}(-A/2)) = \frac{2}{\left(n-1\right)^{(n+1)/(n-1)}(n+1)} \cdot \frac{1}{A^{(n+1)/(n-1)}},
\end{equation}
here we used the superpotential \eqref{eq:superpotential_for_m_equals_1}, and calculated $\varphi_{\rm K}^{}(-A/2)$ using the asymptotics \eqref{eq:kink_for_m_equals_1_asymptotics}.

Within our approach, the force acting on the kink (antikink) can be estimated as
\begin{equation}
    F(A) = - \frac{\partial E}{\partial(2A)} = - \frac{1}{2} \frac{\partial E_{\rm quad}^{}}{\partial A} + \frac{\partial E_{\rm tail}^{}}{\partial A}.
\end{equation}
Differentiating \eqref{eq:E_quad} and \eqref{eq:E_tail}, we get
\begin{equation}\label{eq:force_m_eq_1}
    F(A) = \frac{1}{2} \cdot \frac{n+1}{n-1} \cdot \left[ \varepsilon(n) - \frac{4}{n+1}\left(n-1\right)^{(n+1)/(1-n)} \right] \cdot \frac{1}{A^{2n/(n-1)}}.
\end{equation}
For $n=2$ this formula gives
\begin{equation}\label{eq:force_m_eq_1_n_eq_2}
    F(A) = - \frac{22}{105} \frac{1}{A^4} \approx - \frac{0.21}{A^4},
\end{equation}
and for $n=3$ we obtain
\begin{equation}\label{eq:force_m_eq_1_n_eq_3}
    F(A) = - \frac{3457}{48\:048} \frac{1}{A^3} \approx - \frac{0.072}{A^3}.
\end{equation}
The force is negative, which means attraction between kink and antikink. As $n\to\infty$, the degree of $A$ in the denominator tends to 2 from above, while the numerical coefficient tends to zero.

In the paper \cite{Christov.PRL.2019} for a model with potential
\begin{equation}\label{eq:Christov_potential}
    V(\varphi)=\frac{1}{2}\left(1-\varphi^2\right)^2\varphi^{2n}
\end{equation}
the following asymptotic estimate of the kink-antikink force was obtained in the case of overlapping power-law tails (near the vacuum $\bar{\varphi}=0$):
\begin{equation}\label{eq:Christov_force}
    F(A) \approx - \displaystyle\frac{1}{2}\left[\frac{-\sqrt{\pi}\,\Gamma(\frac{n-1}{2n})}{\Gamma(-\frac{1}{2n})}\right]^{\frac{2n}{{n-1}}} \frac{1}{A^{2n/(n-1)}}.
\end{equation}
The authors of \cite{Christov.PRL.2019} used an adiabatic ansatz for accelerating kinks and reduced the problem to integrating a first-order ordinary differential equation -- a version of the Bogomolny equation \cite{BPS1,BPS2}. (Note that in \cite{Manton.JPA.2019} the same method was independently applied to kinks of the $\varphi^8$ model, and chronologically published even somewhat earlier.)

It is noteworthy that the degree of $A$ is the same in both estimates \eqref{eq:force_m_eq_1} and \eqref{eq:Christov_force}, while the numerical coefficients are expectedly different. The point is that potentials \eqref{eq:potential_general} and \eqref{eq:Christov_potential} differ significantly. In particular, there are three vacua in the model \eqref{eq:Christov_potential}, while the model \eqref{eq:potential_general} has only two vacua. The formula \eqref{eq:Christov_force} gives for $n=2$
\begin{equation}
    F(A) \approx -1.4771/A^4,
\end{equation}
and for $n=3$
\begin{equation}
    F(A) \approx -0.1723/A^3,
\end{equation}
which should be compared with \eqref{eq:force_m_eq_1_n_eq_2} and \eqref{eq:force_m_eq_1_n_eq_3}. For large $n$, the ratio of the coefficients in \eqref{eq:force_m_eq_1} and \eqref{eq:Christov_force} is close to one.

In the work \cite{Manton.JPA.2019} for the potential \eqref{eq:Christov_potential} with $n=2$ (model $\varphi^8$), N.~S.~Manton estimated the kink-antikink force as
\begin{equation}\label{eq:Manton_force}
    F(A) \approx - \frac{88}{105} \frac{1}{A^4} \approx - \frac{0.84}{A^4}.
\end{equation}
Using a method similar to those developed in \cite{Christov.PRL.2019} and \cite{Manton.JPA.2019}, P.~d'Ornellas in the paper \cite{dOrnellas.JPC.2020} for the same potential \eqref{eq:Christov_potential} with $n=2$ obtained the estimate
\begin{equation}\label{eq:dOrnellas_force}
    F(A) \approx - \frac{\left(\Gamma\left(\frac{1}{4}\right)\right)^4}{2048\:\pi^2} \frac{1}{A^4}\approx - \frac{1.48}{A^4}.
\end{equation}
Some of the papers mentioned above also compare analytical estimates with experimental data, demonstrating good agreement.

\section{Numerical simulations}
\label{sec:Numerical}

This section presents the results of numerical experiments that allowed us to evaluate the force of interaction between kink and antikink and compare it with theoretical prediction \eqref{eq:force_m_eq_1}.

\begin{figure}[t!]
    \centering
    \subfigure[\:$n=2$]{
    \includegraphics[width=0.48\linewidth]{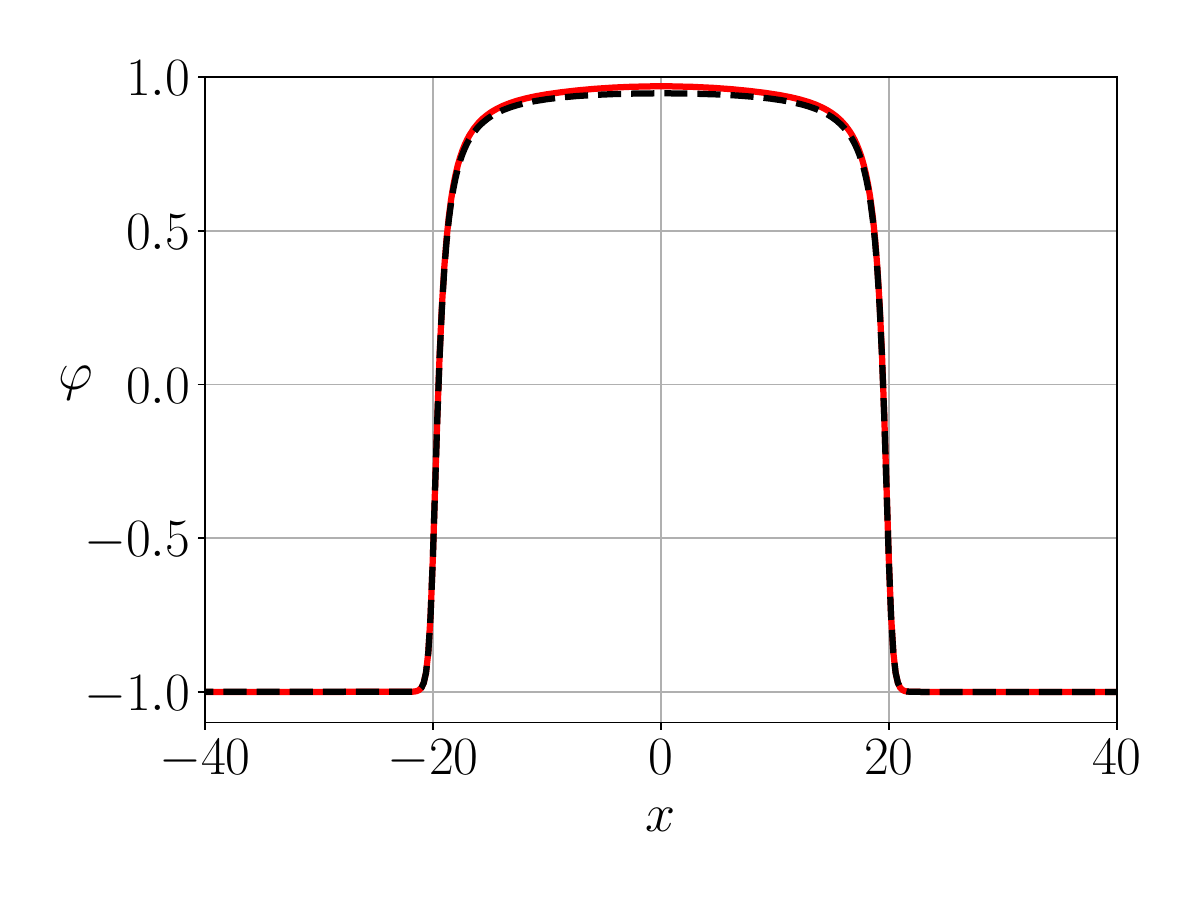}
    }
    \subfigure[\:$n=3$]{
    \includegraphics[width=0.48\linewidth]{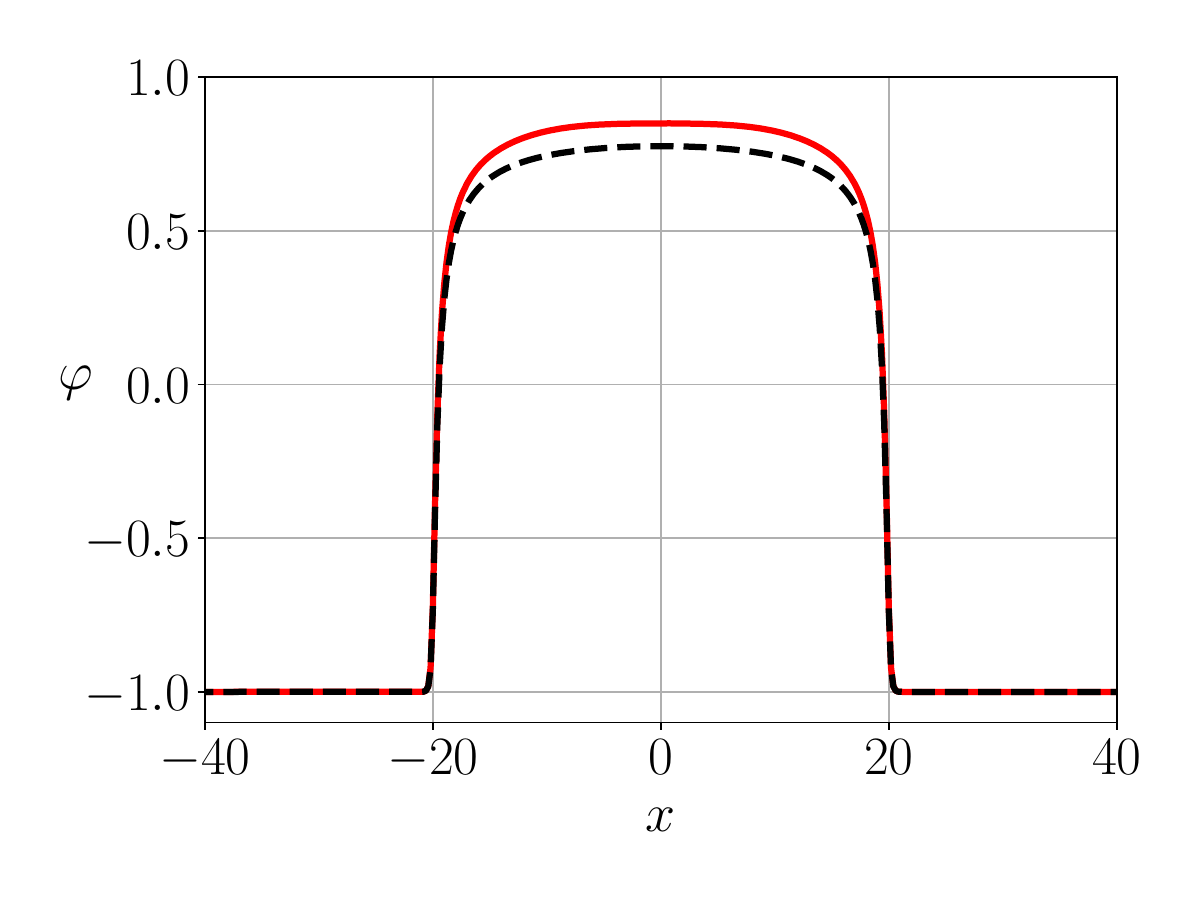}
    }
    \caption{The results of minimization of the functional \eqref{eq:fun-minimize}. Red solid curves show the minimized ansatz, while black dashed curves show the initial approximation~\eqref{eq:zero-approx}.  The half-distance is $A=20$.}
    \label{fig:ini-anzats}
\end{figure}

As an initial condition, we used the configuration in the form of static kink and antikink located at points $x=-A$ and $x=A$, respectively. Since in the case of power-law asymptotics ordinary ansatzes are not applicable \cite{Christov.PRL.2019,Christov.PRD.2019,Christov.CNSNS.2021}, we used a static ansatz, modified in such a way as to minimize the functional
\begin{equation}\label{eq:fun-minimize}
\begin{cases}
    S[\varphi(x)] = |\varphi''(x) - dV/d\varphi|^2,\\
    \varphi(-A) = \varphi(A) = \varphi^{(0)},
\end{cases}
\end{equation}
where $\varphi^{(0)}$ is the value of the field obtained from \eqref{eq:kink_for_m_equals_1} at $x-x_0=0$. We minimized the functional \eqref{eq:fun-minimize} numerically, see Appendix~\ref{ap:num-details} for details. As a zero approximation we took the following kink-antikink configuration:
\begin{equation}\label{eq:zero-approx}
    \varphi(x) = \frac{1}{2}
    \left[\varphi_\mathrm{K}^{}(x+A)+1\right]\left[\varphi_{\Bar{\mathrm{K}}}^{}(x-A)+1\right] - 1.
\end{equation}
The results of the minimization of the functional \eqref{eq:fun-minimize} --- the minimized ansatzes --- together with the initial configurations \eqref{eq:zero-approx} are shown in Fig.~\ref{fig:ini-anzats}.

\begin{figure}[t!]
    \centering
    \subfigure[\:$n=2$]{\includegraphics[width=0.48\linewidth]{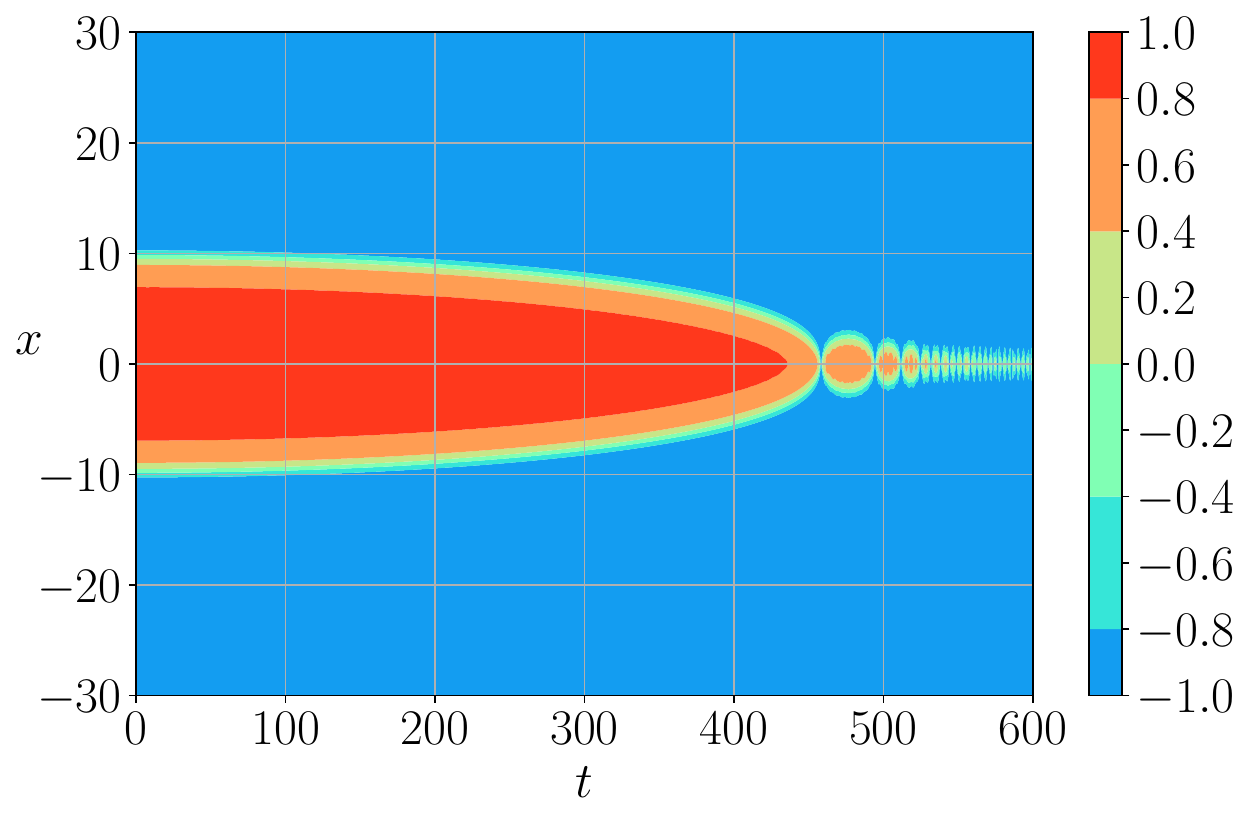}}
    \hfill
    \subfigure[\:$n=3$]{\includegraphics[width=0.48\linewidth]{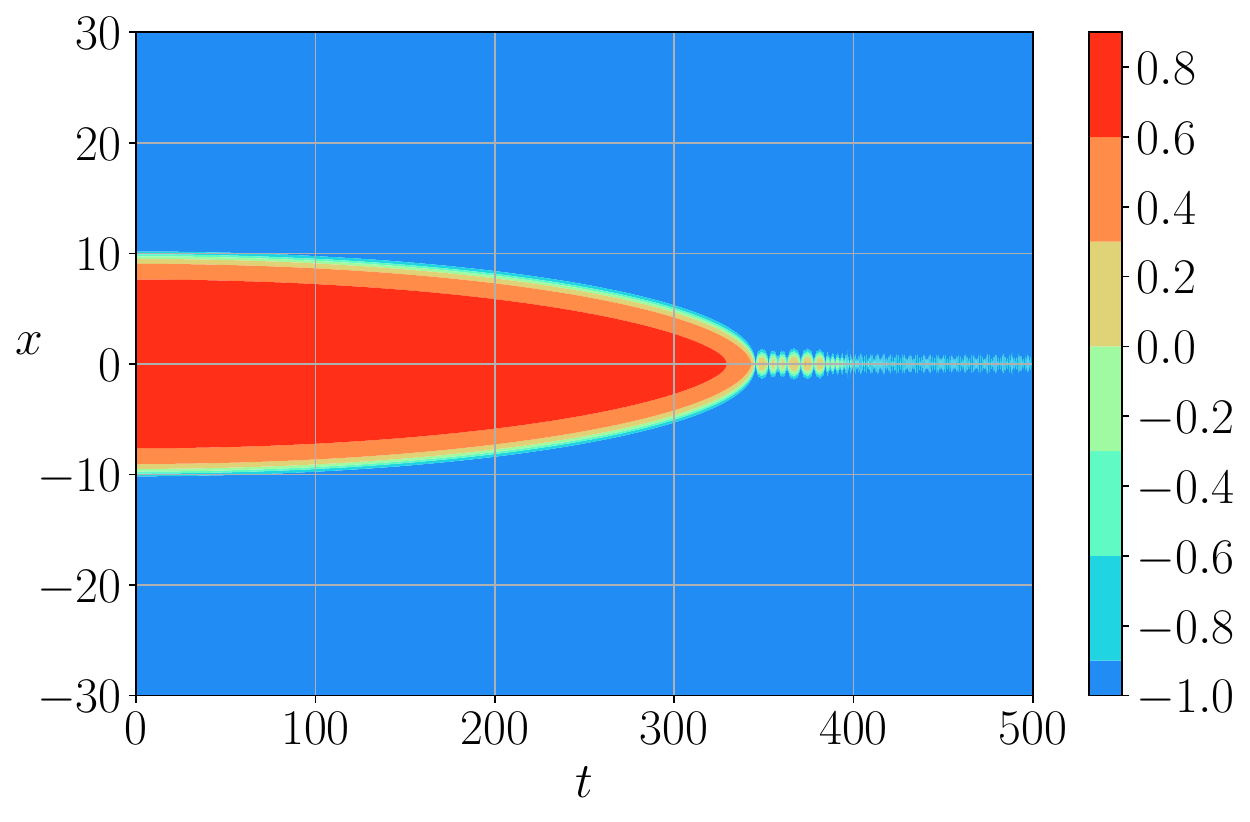}}
    \centering
    \caption{The results of the numerical solution of the equation of motion with the initial condition in the form of the minimized ansatz at $A=10$ and zero initial velocity.}
    \label{fig:eqmo-solu-3d}
\end{figure}

Then, the evolution of the minimized ansatz in accordance with the equation of motion \eqref{eq:eqmo} was numerically found using an explicit second-order finite-difference method.
A typical picture is shown in Fig.~\ref{fig:eqmo-solu-3d}. The initially resting kink and antikink begin to approach each other due to mutual attraction and eventually collide and form a decaying bound state. It can be seen that at $n=3$ it takes less time for the kinks to collide, which indicates a greater force of attraction than at $n=2$.

\begin{figure}[t!]
    \centering
    \subfigure[\:$n=2$]{\includegraphics[width=0.48\linewidth]{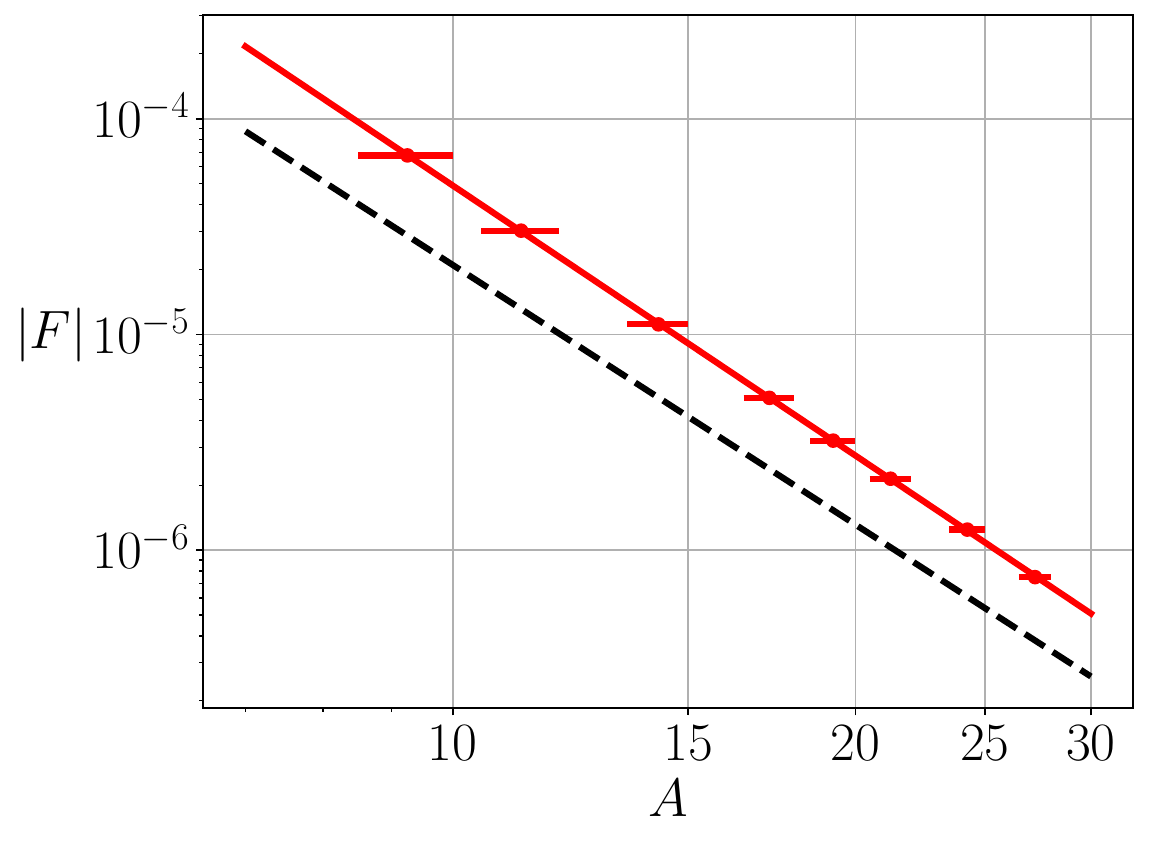}}
    \hfill
    \subfigure[\:$n=3$]{\includegraphics[width=0.49\linewidth]{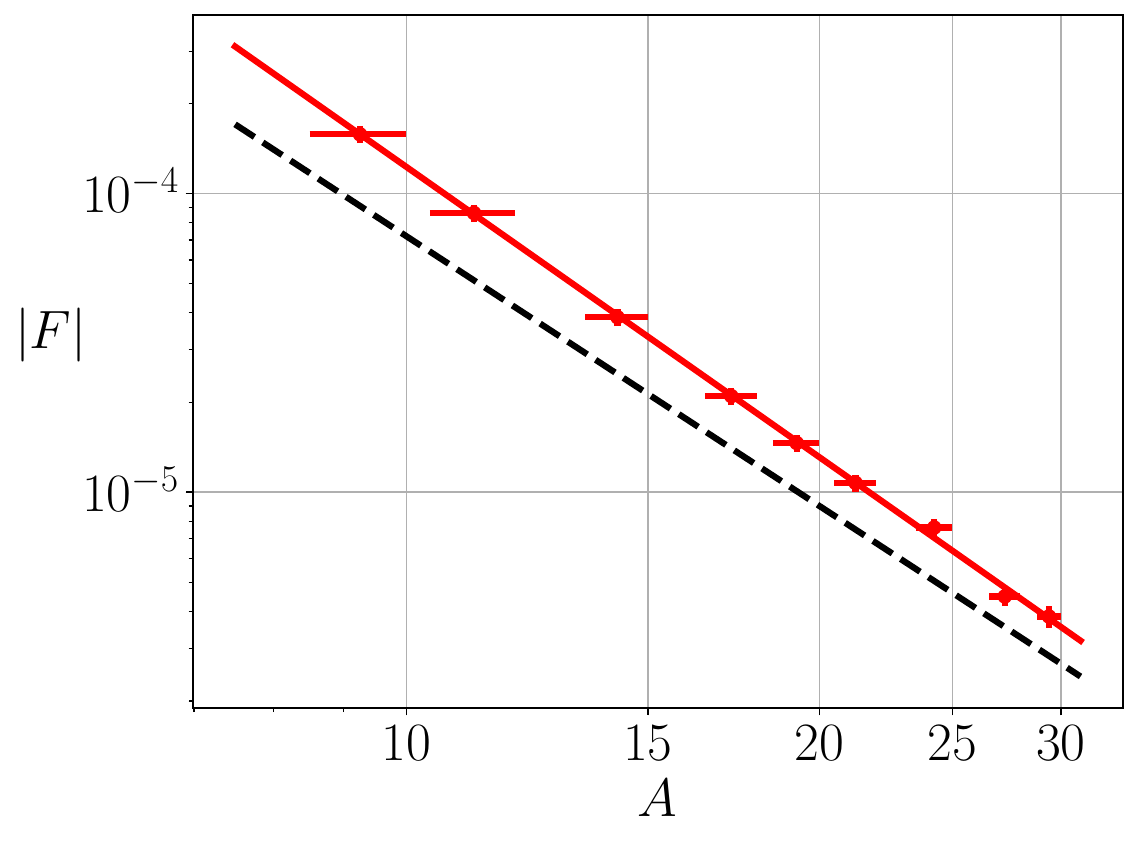}}
    \caption{Comparison of numerical (red dots with error bars and interpolating straight lines) and analytical (black dashed lines, Eqs.~\eqref{eq:force_m_eq_1_n_eq_2} and \eqref{eq:force_m_eq_1_n_eq_3}) results on the kink-antikink force. The following parameters are used for the red lines: $C_2^{}=0.71$, $m_2^{}=4.16$, $C_3^{}=0.21$, $m_3^{}=3.23$.}
    \label{fig:forces-numerical}
\end{figure}

The analysis of the kinks motion under the influence of mutual attraction allowed us to obtain numerical (experimental) values of the force at various $A$. The experimental points together with the theoretical curves are shown in Fig.~\ref{fig:forces-numerical}.
It is seen that for sufficiently large $A$ the agreement is good. Analysis of the data shown in Fig.~\ref{fig:forces-numerical} allows to obtain the following experimental fits for $n=2$ and $n=3$:
\begin{equation}\label{eq:force_fit_n_eq_2}
    F_2^{}(A) = - \frac{C_2^{}}{A^{m_2^{}}}
\end{equation}
with $\ln C_2 = -0.34\pm 0.15$ and $m_2^{}=4.16\pm 0.05$;
\begin{equation}\label{eq:force_fit_n_eq_3}
    F_3^{}(A) = - \frac{C_3^{}}{A^{m_3^{}}}
\end{equation}
with $\ln C_3 = -1.57\pm 0.11$ and $m_3^{}=3.23\pm 0.04$.

\section{Conclusion}
\label{sec:Conclusion}

We have obtained asymptotic estimates of the kink-antikink forces in a family of $\varphi^{2n+2}$ models with polynomial potentials of the form $V(\varphi) = \frac{1}{2}\left(1+\varphi\right)^{ 2}\left(1-\varphi\right)^{2n}$, where $n\ge 2$ is integer. A distinctive feature of kink solutions in these models is that they have one exponential asymptotics (as the field approaches vacuum $\bar{\varphi}_1^{}=-1$) and one power-law asymptotics (as the field approaches vacuum $ \bar{\varphi}_2^{}=1$). We were interested in the case when the power-law tails overlap, and hence long-range interaction appears between the solitons, which manifests itself in the power-law decrease of the force with distance.

By using a quadratic function to approximate the field in the region of the power-law tails overlap, we obtained the dependence of the kink-antikink force in the case of large distance. We have shown that the force decreases as the $\frac{2n}{n-1}$-th power of the (half)distance. We also performed numerical simulations of the kink-antikink interaction and obtained experimental values of the force. Comparison with asymptotic estimates shows good agreement, especially for sufficiently large distances between kinks.

As for the case when the exponential tails overlap, i.e., when the kinks face each other with their exponential asymptotics, it is well studied in the literature, see, e.g., Sections 5.2 and 5.3 in \cite{Manton.book.2004}. We have carried out numerical experiments to find the interaction force and also obtained a theoretical asymptotic estimate of this force. The results show that the interaction force decreases exponentially and can be easily found by the method described in \cite{Manton.book.2004}.

In the paper \cite{Blinov.CSF.2022}, kinks with both left and right power-law asymptotics were obtained in models with potentials $V(\varphi) = \frac{1}{2}\left(1+\varphi\right)^{ 2m}\left(1-\varphi\right)^{2n}$, where $m\ge 2$, $n\ge 2$ are integers:
\begin{equation}\nonumber
    x - x_0^{} = -\frac{1}{(1-\varphi)^{n-1}}\sum\limits_{j=1}^{m-1} \frac{C_{m+n-2}^{j-1}}{2^j\left(m-j\right)C_{m-1}^{j-1}} \cdot \frac{1}{(1+\varphi)^{m-j}} +
\end{equation}
\begin{equation}\label{eq:kink_general}
    + \frac{C_{m+n-2}^{m-1}}{2^{m-1}} \sum\limits_{l=1}^{n-1}  \frac{1}{2^l\left(n-l\right)} \cdot\frac{1}{(1-\varphi)^{n-l}} + \frac{C_{m+n-2}^{m-1}}{2^{m+n-1}}\ln\frac{1+\varphi}{1-\varphi}.
\end{equation}
If we impose a condition $\varphi=0$ for $x=0$, then the constant $x_0^{}$ in Eq.~\eqref{eq:kink_general} is $x_0^{}=\displaystyle\sum\limits_{j=1}^{m-1}\frac{C_{m+n-2}^{j-1}}{2^j\left(m-j\right)C_{m-1}^{j-1}}-\frac{C_{m+n-2}^{m-1}}{2^{m-1}}\sum\limits_{l=1}^{n-1}\frac{1}{2^l(n-l)}$. Calculating forces between kinks in these models could be the next step. For example, interesting question to be resolved is whether the non-overlapping power-law tails make any contribution to the force.

The study of resonance phenomena in collisions of kinks with power-law asymptotics obtained in \cite{Blinov.CSF.2022}, as well as establishing connection with the excitation spectrum of the ``kink+antikink'' configuration, is certainly also of interest. Such a study raises a number of interesting questions. For example, in \cite{Belendryasova.CNSNS.2019} an ansatz in the form of simple sum of kink and antikink was used to construct the stability potential of the ``kink+antikink'' configuration, which defines the excitation spectrum of the configuration. In light of the successful application of the minimization algorithm, it may be more efficient to apply the minimized ansatz to construct the stability potential.

\section*{Acknowledgments}

This work was partly funded by the Ministry of Science and Higher Education of the Russian Federation, Project ``New Phenomena in Particle Physics and the Early Universe'' FSWU-2023-0073.

\appendix
\section{Numerical details}
\label{ap:num-details}

\subsection{Initial conditions for numerical study of the kink-antikink interaction}

\def\vecp{\pmb{\varphi}}
\def\mat{\pmb{\mathrm{D}}}
\def\rhs{\pmb{\mathrm{V_\varphi}}}
\def\modrhs{\pmb{\Hat{\mathrm{V}}_\varphi}}

We are interested in a configuration (ansatz) of the type of static kink and antikink at the points $x=\mp A$ that minimizes the functional \eqref{eq:fun-minimize}. We build a uniform regular grid on the domain $[-x_b^{},x_b^{}]$, with the boundary $x_b^{}$ being sufficiently distant from the position of the kink center, e.g., $x_b^{}=2A$. As a result, we have an $N$-component vector $\pmb{x} = \{x_0^{}, \dots, x_{N-1}^{}\}$, where $h_x^{} = x_{i+1}^{} - x_i^{}$ is the spatial step. We will search for the minimum of the functional \eqref{eq:fun-minimize} numerically on the grid $\pmb{x}$. Consider a grid function $\vecp$ such that $\varphi_i^{} = \varphi(x_i^{})$. We approximate the second derivative at each point $x_k^{}$ with the second order of accuracy as
\begin{equation}\label{eq:2nd-der-approx}
    \varphi_k'' \approx \frac{\varphi_{k+1}^{} - 2\varphi_{k}^{} + \varphi_{k-1}^{}}{h_x^2}.
\end{equation}
At the boundary of the domain, i.e., at the points $x_0^{}$ and $x_{N-1}^{}$, we set
\begin{gather}
    \varphi'_0 \approx \frac{\varphi_{1}^{} - \varphi_{0}^{}}{h_x^{}} = 0,\\
    \varphi'_{N-1} \approx \frac{\varphi_{N-1}^{} - \varphi_{N-2}^{}}{h_x^{}} = 0.
\end{gather}
Let's transform the second derivative of the grid function to vector form:
\begin{equation}
    \left\{\varphi''\right\}_{k=0}^{N-1} = \mat \vecp,
\end{equation}
where $\mat$ is a tridiagonal matrix with non-zero components
\begin{gather}
    \mat_{k,k-1}^{} = 1/h_x^2,\: \mat_{k,k+1}^{} = 1/h_x^2,\: \mat_{k,k}^{} = -2/h_x^2, \quad k=1,\dots,N-2, \\
    \mat_{0,0}^{} = -1/h_x^{},\: \mat_{0,1}^{} = 1/h_x^{}, \\
    \mat_{N-1,N-2}^{} =-1/h_x^{},\: \mat_{N-1,N-1}^{} =1/h_x^{}.
\end{gather}
We approximate the expression $\varphi'' - dV/d\varphi$ using a grid function and reduce it to vector form:
\begin{equation}
    \varphi'' - \frac{dV}{d\varphi} \approx \mat\vecp - \rhs,
\end{equation}
where the vector $(\rhs)_k^{} = \displaystyle\frac{dV}{d\varphi}(\varphi_k^{})$ for $k=1,\dots,N-2$, and $(\rhs)_0^{} = (\rhs)_{N-1}^{} = 0$ at the boundaries.

The condition $\varphi(-A) = \varphi(A) = \varphi^{(0)}$ in \eqref{eq:fun-minimize} means the requirement to fix the centers of the kink and antikink at the grid points with numbers $i_{-A}^{}$ and $i_{A}^{}$ such that $x_{i_{-A}^{}}^{}=-A$ and $x_{i_{A}^{}}^{}=A$. Taking this term into account changes matrix $\mat$ into matrix $\Hat{\mat}$ as follows:
\begin{gather}
    \Hat{\mat}_{i,k} = \mat_{i,k}, \quad i\neq  i_{-A}^{}, \\
    \Hat{\mat}_{i_{-A}^{}, k} = 
    \begin{cases}
        0,\quad k\neq i_{-A}^{},\\
        1,\quad k=i_{-A}^{},
    \end{cases}\\
    \Hat{\mat}_{i_{A}^{}, k} = 
    \begin{cases}
        0,\quad k\neq i_{A}^{},\\
        1,\quad k=i_{A}^{}.
    \end{cases}
\end{gather}
The vector $\rhs$ also changes into $\modrhs$ as
\begin{gather}
    (\modrhs)_{i} = (\rhs)_{i} , \quad i\neq  i_{-A}^{}, \\
    (\modrhs)_{i_{-A}^{}} = \varphi^{(0)},\\
    (\modrhs)_{i_{A}^{}} = \varphi^{(0)}.
\end{gather}
Thus, the discrete form of the functional~\eqref{eq:fun-minimize} looks like
\begin{equation}\label{eq:discr-fun-min}
    S = (\Hat{\mat}\vecp - \modrhs)^T(\Hat{\mat}\vecp - \modrhs).
\end{equation}
Next, we used an initial approximation in the form of~\eqref{eq:zero-approx} and used the gradient descent method to find its local minimum. The accuracy of the gradient descent method is $10^{-5}$, the grid step used in the calculations is $h_x^{}=0.1$.

\subsection{Numerical simulation of the kink-antikink dynamics}

We also used the matrix of derivatives $\mat$ introduced in the previous subsection to approximate the second derivative with respect to the coordinate in the equation of motion. After reducing the second derivative to vector form for the grid function $\vecp$, the equation of motion was reduced to a Cauchy problem with the initial condition in the form of the minimized ansatz and the zero first derivative with respect to time. The Cauchy problem was solved using an explicit finite-difference scheme of the second order in time:
\begin{gather}
    \vecp^{m+1} = 2\vecp^{m} - \vecp^{m-1}
    + h_t^2\mat \vecp^{m} - h_t^2\rhs^{m}, \\
    \vecp(t=0) = \vecp^{0}, 
\end{gather}
where $m$ denotes the $m$-th time layer $t=t^m$, $\rhs^m = \rhs(\vecp^m)$, and $\vecp^0$ is the value of the field at $t=0$ (the minimized ansatz). To start the calculation, we also need the value  of the field at the first moment of time $t=h_t^{}$:
\begin{gather}    
    \vecp(t=h_t^{})\approx \vecp(0)+h_t\vecp_t'(0)+\frac{h_t^2}{2}\vecp_{tt}''(0),
\end{gather}
but since $\vecp_t'(0)=0$, and the second derivative with respect to time is expressed using the equation of motion $\vecp_{tt}''(0)=\mat\vecp(0)-\rhs$, then the field at $t=h_t^{}$ is
\begin{gather}
    \vecp^{1} = \vecp^{0} + \frac{h_t^2}{2}\left(
    \mat \vecp^{0} - \rhs^0
    \right).
\end{gather}
In the calculations, the steps $h_x^{}=0.1$ and $h_t^{}=0.01$ were used, for which the Courant condition $h_t^{}/h_x^{} < 1$ is satisfied.

\subsection{Numerical evaluation of the kink-antikink attraction force}

To numerically find the attractive force of the kink and antikink, we used Newton's second law:
\begin{equation}
    F = M_\mathrm{K}^{}\cdot (a\pm\Delta a),
\end{equation}
where $M_\mathrm{K}^{}$ is the kink mass \eqref{eq:mass_for_m_equals_1}, $a$ is the kink's acceleration, and $\Delta a$ is the acceleration calculation error.

Let the kink initially located at the point $x=-A$ has traveled a distance $s\ll A$ in time $t_s^{}$. Then its acceleration can be estimated as
\begin{equation}
    a = \frac{2s}{t^2_s}, 
\end{equation}
The error in calculating acceleration is
\begin{equation}
    \Delta a = 
\sqrt{\left(
\frac{\partial a}{\partial s}
\right)^2 h_x^2 +
\left(
\frac{\partial a}{\partial t}
\right)^2 h_t^2} \approx \frac{2 h_x}{t^2_s}.
\end{equation}
The second term under the square root is much smaller than the first, since the interaction force is very small, and hence $s/t_s^{}\ll 1$. The steps $h_x^{}$ and $h_t^{}$ are taken from the scheme for solving the equation of motion, i.e., $h_x=0.1$ and $h_t=0.01$.

The position of the kink center was determined as the position of the maximum energy density~\eqref{eq:energy} of the moving kink. Since the kink moves, the half-distance corresponding to the force $F$ was taken as the average $A-s/2$ with an uncertainty of $\pm s/2$.

\end{document}